\documentclass[12pt,a4paper]{article}
\usepackage[koi8-r]{inputenc}
\usepackage[russian]{babel}
\usepackage{epsfig}
\usepackage{psfig}
\usepackage{graphicx}

\begin{document}

\sloppy

\thispagestyle{empty}

\begin{center}

{\LARGE\bf{The dispersion formula and the light group velocity in a water}} 
\end{center}
\begin{center}
\author\footnotesize{\bf{$I.A. Danilchenko$}}\\
Institute for Nuclear Research, Russian Academy of Science(Moscow),
	  Russian Federation,\\
Corresponding E-mail: iad@pcbai10.inr.ruhep.ru\\
\end{center}

{\it\small {The parametric formulas are obtained whereby the refractive index and 
the light group velocity could be calculated for any water state 
(salinity, pressure, temperature) inside of the light
wavelength region
from 200nm to 1200nm. These formulas are suitable to be used in
underwater neutrino telescope activity.}}
\begin{center}
		{\bf Introduction}
\end{center}
			
        In underwater Cherenkov neutrino telescopes /1$\div$5/ water acts 
	as a radiator of Cherenkov photons generated by charge particles.
	Cherenkov photons are emitted in cone with half-angle defined by the
	mass and momentum of the particle and the refractive index of the
	medium (water in our case)
	$$cos\Theta=1/(\beta\cdot n(\lambda)),\eqno (1)$$    
        where $\beta$ is velocity of particle, $n(\lambda)$ is 
	refractive index and $\lambda$ is wavelength.\
	
	Because the medium (water) 
	is disperseve, the group velocity 
	(velocity of light signal transmission in transparent material) 
	is
	different from phase velocity \\
	$v_{ph}(\lambda)=c/n(\lambda$) and
	being characterised by a group index of this medium
	$$v_{gr}(\lambda)=c/n_{gr}(\lambda),\eqno(2)$$
	$$n_{gr}(\lambda)=n(\lambda)-\lambda\cdot (dn/d\lambda), \eqno (3)$$\	
	where $c$ is the light velocity in vacuum.\
		
	 This reality is taken into account for example in construction
	of a large-area Time Of Flight(TOF) counters. The group index of
	scintillator is $n_{gr}=1.75$ compared with refractive index
	$n=1.5$ /6/. In the DUMAND-type installation activity  a phase 
	velocity was used for estimation of a photon TOF till it was 
	demonstrated /7/ that such approach leads to errors in results
	of TOF-calculations. For photon passed through
	100m water layer these errors  vary from 10ns to 18ns when
	the light wavelength varies from 500nm to 350nm. A disregard of the 
	$v_{gr}-v_{ph}$ difference does not affect errors in track
	reconstruction for the existing projects /8/, but, as
	Baikal Neutrino Telescope experience shows, only $v_{gr}$
	must be used for time-calibration of the detector modules 
	with the help of an outside laser  and its value
	have to be known with a quite high accuracy.\
	
	The measurement of the group velocity may be performed
	with a help of detector modules lighted by an outside 
	laser which shines through water /9/. Comparison of 
	the experimental result with the value of this physical 
	constant gives
	an unique possibility to verify the accuracy and reliability
	of all time measuring units and all time calibration procedures.
	To prove the height level of the TOF measuring the accuracy of 
	$v_{gr}$ calculation have to be higher than this level and
	therefore all factors which affect the $n_{gr}$ value 
	(wavelength dependence, temperature, pressure, salinity)
	must be taken into account.
	  
\begin{center}
	{\bf 1.  Dispersion formulas for pure and sea water}
\end{center}
	
	In applied optics a calculation of a refraction index for any
	wavelength in the restricted wavelength region is performed 
	using so called dispersion formulas. These formulas have the same
	structure for different class of material but differ
	usually in the formula coefficients. Thus for example for optical
	colourless glass in the wavelength interval 
	365$\div$10139 nm the following formula is used /10/:
	$$n^{2}=A_{1}+A_{2}\cdot \lambda^{2}+A_3\cdot \lambda^{-2}
	+A_4\cdot \lambda^{-4}+A_5\cdot \lambda^{-6}+
	A_6\cdot \lambda^{-8} ,\eqno (4)$$\
	where $A_i$ are the coefficients of the dispersion formula, which
	depend on the glass class and each class is characterised by
	one's own set of $A_i$ /10/.\

	There is abundance of table data/10-12/ which contain the results 
	of the refraction index measurements performed for 
	pure (salinity S=0 $^0/_{00}$)
	water, atmospheric pressure $P=P_{atm}$ and 
	temperature $t=20^0 C$. The formula (5) represent the best
	fit to these data and the lower solid line 1 on Fig.1 shows
	the result of the refraction index calculation by means of
	this expression. The points on this 
	line correspond to table data used in the fitting procedure.
$$n_{0}^2(\lambda)=1.7527-2.55\cdot 10^{-3}\cdot \lambda^4-
5.0\cdot 10^{-3}\cdot \lambda^2+9.9\cdot 10^{-3}\cdot \lambda^{-2}-$$
$$-4.0\cdot 10^{-4}\cdot \lambda^{-4}+2.9\cdot 10^{-5}\cdot \lambda^{-6}-
5.0\cdot 10^{-7}\cdot\lambda^{-8},\eqno(5)$$\
where $ \lambda$ in mkm and index '0' means distilled water with t=20$^o$C
and P=Patm. By comparison with expression (4) the term
$A\cdot \lambda^4$ was added to extend the region in which 
dispersion formula is correct since 
$\lambda$=214nm up to $\lambda$=1256nm.\

    Salinity, pressure and temperature 
    affect the value of the water refraction index. The sets of 
    the same black points which are also shown in Fig.1 correspond 
    to the sets of the table data /12/ for different states (S, P, t) of 
    sea water. The dependencies n versus $\lambda$
    for these sets are described by functions 
    $$n(\lambda, S, P, t)=n_0(\lambda, S_0, P_0, t_0)+\bigtriangleup n(S,P,t) \eqno(6)$$
    with quite high precision. The last fact permits to conclude, that
    in distinct with an optical glass, the dispersion formula 
    coefficients are the same for distilled water, sea water and compressed 
    water and they do not depend on temperature. Thus to calculate a refraction
    index of any water for any wavelength the dispersion formula (5) may be
    used as an universal one and just S, P, t -corrections ($\bigtriangleup
    n(S,P,t$)) must be taken into account  
    which do not depend on wavelength.

\newpage
        
\begin{figure}[ht]

\mbox{\epsfig{ file=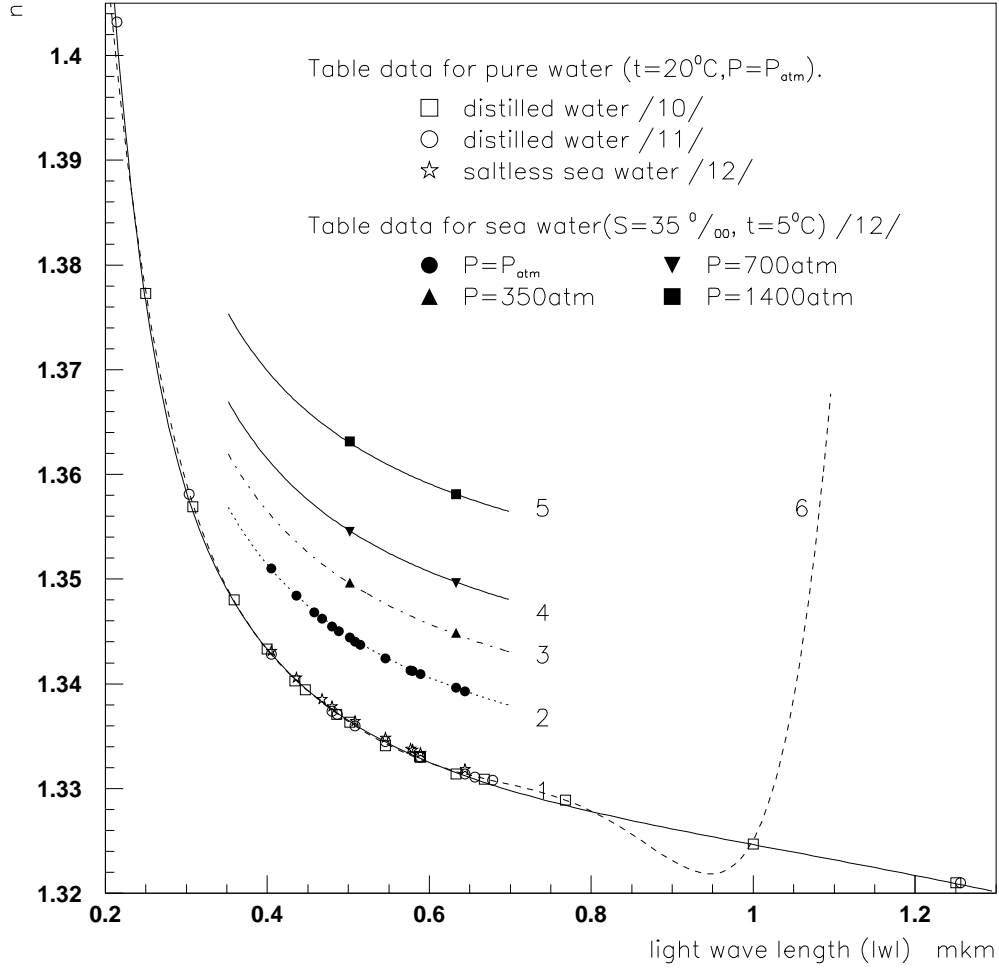, width=15cm, height=15cm }}
       \caption[1]{\small Dispersion formulas and table data.     
       The lines (1-5) were calculated by 
       means of the dispersion formula(5) and line 6 by means polinomial /14/.
1 :  n=n$_0$; 2 : n=n$_0$+0.0081; \\3 : n=n$_0$+0.0132; 
4 : n=n$_0$+0.0182; 5 : n=n$_0$+0.0266.}
\end{figure}
\begin{center}
{\bf2.  S, P, t -corrections of water refractive index}\
\end{center}
    In the simplest approach /7, 8/ the Lorentz-Lorentz formula may be used
    to get such corrections:
    $$\frac{n^2-1}{n^2+2}\cdot \frac{1}{\rho}=\frac{4}{3}\cdot
    \frac{N_A}{M}\cdot\alpha_e  ,\eqno(7)$$ 
where $M$ is the molecular weight, $N_A$ \ - Avogadro 
number, $\alpha_e$ \ - polarizability and 
$$R_{\rho}=\frac{n^2-1}{n^2+2}\cdot \frac{1}{\rho} \eqno(8)$$  
is the specific refractive index. 
If we suppose that the right part of the expression (7) is constant, the
change in refractive index with density is given by
     $$\frac{\bigtriangleup n}{\bigtriangleup \rho}=\frac{(n^2+2)
\cdot (n^2-1)}{6\cdot n\cdot \rho}, \eqno(9)$$
     and calculation of $\bigtriangleup\rho(S, P, t)$ is enough
     for the estimation of $\bigtriangleup n(S, P, t)$.\
     
      Analysis of table 
     data/12/ shows some dependence $R_\rho$ versus $t$ and versus $S$ and
     so the described approach becomes not quite correct. Direct comparison
     of the measured increasing $\bigtriangleup$ $n_{exp}$ which
     occurs due to water 
     temperature fall from 20$^0$C to t$^0$C with the 
     value $\bigtriangleup$ $n_{L-L}$ (formula (7)) shows that
     the measured results exceed the estimated values 
     on 18\%, 23\%, 29\% and 
     37\% for temperatures  
     15$^0$C, 10$^0$C, 5$^0$C and 0$^0$C
     correspondingly.        
     Therefore in the given 
     calculations the experimental results are used as fully as 
     possible and the formula (9) may be used, if such results are not 
     available, but only for $\bigtriangleup n_p$ correction, when 
     density is changed directly due to the pressure increasing.
     To calculate the corrections it is supposed that
     $$\bigtriangleup n(S, P, t)=\bigtriangleup n_{St}(S, P_0, t)+
     \bigtriangleup n_p(S, P, t). \eqno(10)$$



\

The refractive index table data which
have been measured for sea water /12/ were used to get some 
mathematical expressions to describe the dependencies $n$ versus 
$S$, $t$ and $P$. The points on Fig. 2 
correspond to sea water refractive index values
measured under atmospheric pressure for photons 
with $\lambda=583.9nm$ /12/ and the lines are calculated using the 
expression (11), which is the best fit to these points:
$$n(S, P_0, t)=1.3340+0.19705\cdot10^{-3} \cdot S-$$
$$-(0.15655+0.010045\cdot S-0.41077\cdot 10^{-4} \cdot S^2) \cdot 10^{-4}\cdot t-$$
$$-(0.17585-0.14300 \cdot 10^{-2} \cdot S+0.12525 \cdot 10^{-4}\cdot S^2)\cdot 10^{-5} \cdot t{^2}. \eqno(11)$$
  
Because of the salinity varies inside the region $(35\pm4)  ^0/_{00}$ 
in projects planned to be installed in a deep sea, the table data /12/ 
are used
which contain refractive index values ($\lambda=501.7nm$)
measured under high pressure in water with $S=35^0/_{00}$. 
 These points for which the sea depth values (h(m)) are used instead
the pressure values are shown on Fig. 3 and dependence of the refractive index 
versus h and versus t is described well by the expression (12) :
$$n(35, h ,t)=1.3448+0.15024\cdot 10^{-5}\cdot h-0.13122
\cdot 10^{-10}\cdot h^2-$$     
$$-(0.50560+0.80127\cdot 10^{-4}\cdot h-0.99739\cdot 10^{-9}
\cdot h^2)\cdot 10^{-4}\cdot t-$$
$$-(1.3118-0.11469\cdot 10^{-3}\cdot h+0.18712\cdot 10^{-8}
\cdot h^2)\cdot 10^{-6}\cdot t^2. \eqno(12)$$
\\
The expressions (11) and (12) may be used to calculate  
(S, t)-correction and P$_h$-correction correspondingly :
$$\bigtriangleup n_{St}=n(S, P_0,t)-n(0,P_0,t_0), \eqno(13)$$ 
$$\bigtriangleup n_p=n(35,P_h,t)-n(35,P_0,t). \eqno(14)$$
A pure water compressibility is higher than the one for
 salt water and, therefore, according (9) one has 
 $\bigtriangleup n_p(0,P,t)>\bigtriangleup n_p(S,P,t)$. To 
account it (in the absence of direct experimental data) the depth
correction for pure water may be calculated as 
$$\bigtriangleup n_p(0,P_h,t)=\bigtriangleup n_p(35,P_h,t)\cdot K_S,\eqno(15)$$

\begin{figure}[ht]
\mbox{\epsfig{file=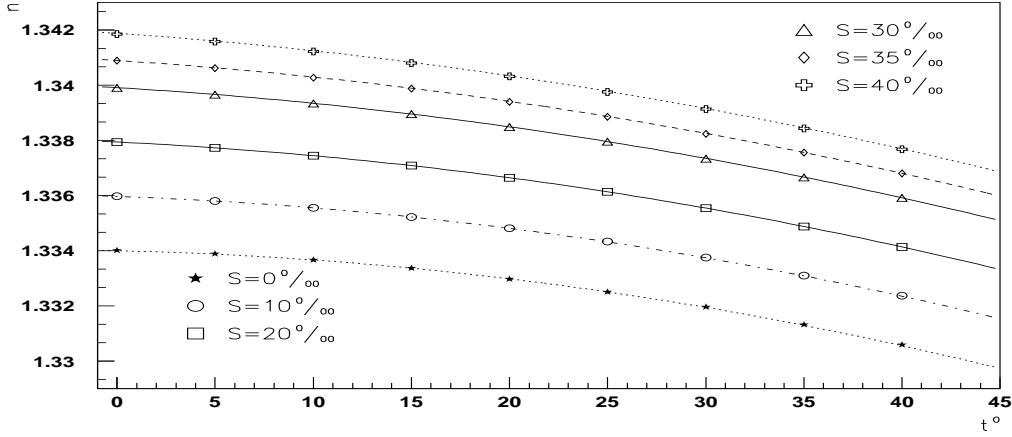,width=15cm,height=7.2cm}}
\caption[1]{\small Refractive index $(\lambda=589.3 nm)$ of sea water under 
atmospheric pressure /12/.
}
\end{figure}
\begin{figure}[h]
\mbox{\epsfig{file=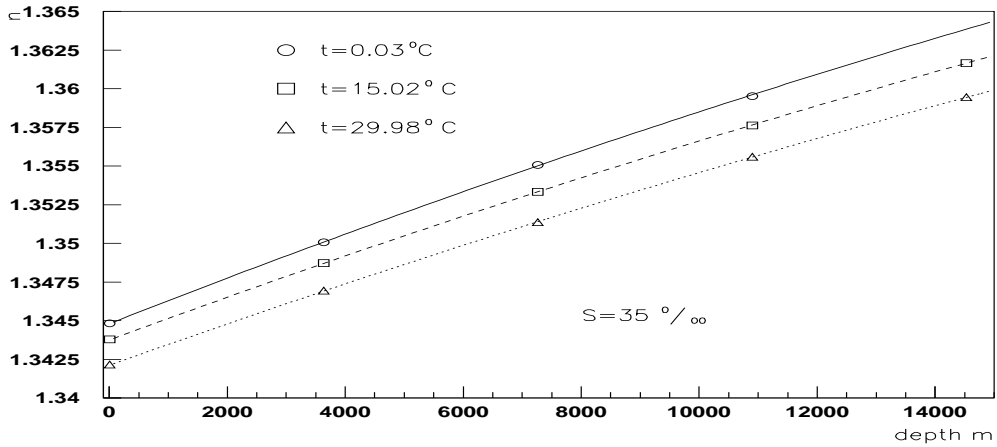,width=15cm,height=7.2cm}}
\caption[1]{\small Refractive index $(\lambda=501.7nm)$ of 
sea water (S=35$^0/_{00}$) for various depths /12/.
} 
\end{figure}
where $K_S=\bigtriangleup\rho(0, P_h, t)/\bigtriangleup\rho(35, P_h, t)$. Analysis of the table data /12/ gives $K_S=1.07\pm 0.01$
for $t=(0\div 5)^o C$ and $h=(0\div 14000)m$.  
\begin{center}
{\bf 3. $\bigtriangleup$n(S, P, t) -corrections and group light 
velocity for existing projects}
\end{center}\

The refractive index corrections caused by environmental 
parameters ($S,P_h, t$) of detector sites, and values of the light group
velocity for $\lambda$=475 nm which have been calculated by means
expressions 5$\div$6 and 10$\div$14 are put in the table. The points on Fig. 4 
reflect the results of such calculations for the various values of
light wavelength.\
 
\begin{center}  
\begin{tabular}{|l|c|c|c|c|c|c|}
\hline
 Detector & S ($^0/_{00}$) & h (m) & t ($^0$C) & $ \bigtriangleup n(S, P_h, t)$ & v$_{gr}$(10$^{10}$cm/s)\\[5pt]
\hline
 DUMAND & 35 & 4600 & 1.3 & 0.0147 & 2.1729\\
 NESTOR & 38.7 & 3800 & 14 & 0.0129 & 2.1757\\
 ANTARES & 33.65 & 2500 & 13.2 & 0.0103 & 2.1799\\
 BAIKAL & 0.12 & 1150 & 3.4 & 0.0027 & 2.1919\\
\hline
\end{tabular}
\end{center}
\begin{figure}[ht]
\mbox{\epsfig{file=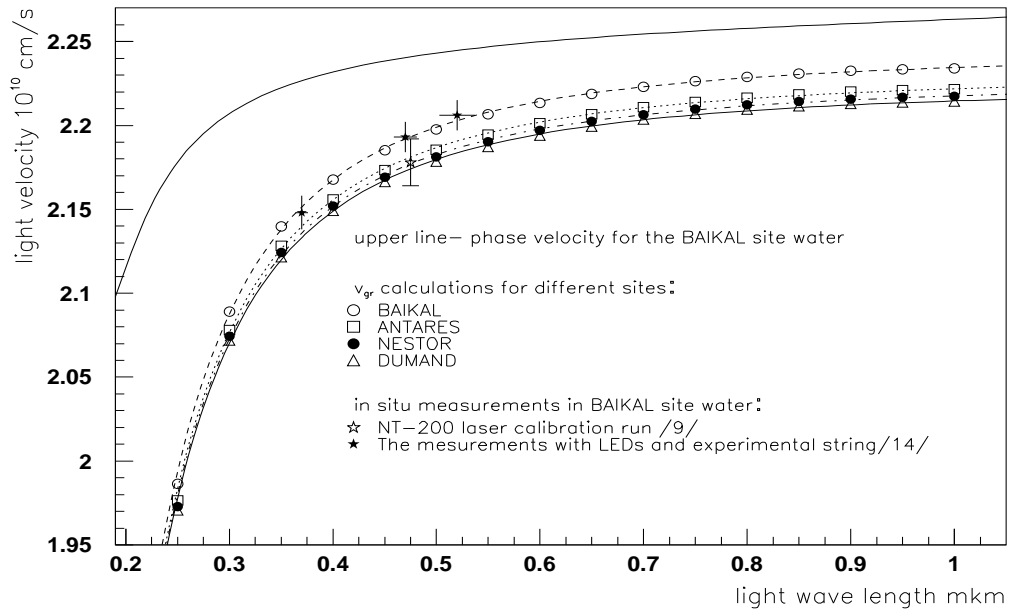,width=15cm,height=10cm}}
\caption[1]{\small The dependencies of calculated light velocity  
versus light wavelength for the existing underwater neutrino telescope
projects. The points and upper line have been calculated 
by means the expressions (2$\div$3), (5$\div$6) and (10$\div$14). The stars 
correspond to
the experimental results/9,14/. The other lines are calculated by means
the formula(16).
}
\end{figure} \

It seems to be completely justified to simplify the calculation procedure
by using the following formula which is the best fit for these points:
$$v_{gr}(\lambda,S,P,t)=c/(1.333+0.57\cdot10^{-2}\cdot\lambda^{-2}
+0.3\cdot10^{-3}\cdot\lambda^{-4}+\bigtriangleup n(S, P, t)). \eqno(16)$$\

\newpage

\begin{center}    
    
{\bf 4. Comparison with the experimental result}\

\end{center}    
       The first direct measurements of the light ($\lambda$=475nm) signal velocity
       ($V$)  have been performed in the water volume of Baikal Neutrino 
      Telescope (NT-200) /9/. These measurements did not
      require any add devices and any changes in NT-200 standard regime. Data 
      of usual laser calibration run has been used to do this. During such
      runs NT-200 Channels (Ch) detect the flashes of laser, which has 
      attached to NT-200 central string on 20m lower than the lowest Ch. Laser 
      generates five flash series with the decreased photon intensity. Only 
      central string data have been taken into account to avoid the influence 
      of the neighbour strings co-ordinates ambiguities. Mean measured 
      values of Time Difference($T_{ij}$) for 
      any of two Ch's (45 pairs in the 
      total) and the (measured before) spaces between Ch's 
      attachment points
      and laser attachment point were used as an input 
      information. Increasing 
      of Ch time delay with the laser flash intensity decreasing 
      (time walk effect) was observed for all Ch's. The special 
      procedure /9/ was 
      applied to compensate an influence of this effect on the final result.
      Analysis of the $V$-distributions measured for 
      different $S_{ij}=S_i-S_j$ 
      ($S_i$ 
      is the distance from laser to Ch$_i$) has demonstrated 
      no dependence of
      mean velocity values $<V>$ on $S_{ij}$ (Fig 5A). For selection 
      criterion $S_i-S_j$ $>$ 12.5 m  
      the expected value of the laser signal velocity 
                    $$V=(2.178 \pm 0.014)\cdot10^{10} cm/s$$ 
      had been obtained using data of 30 pairs of Ch's which satisfy 
      this criterion (Fig 5B).
\begin{figure}[ht]
\mbox{\epsfig{file=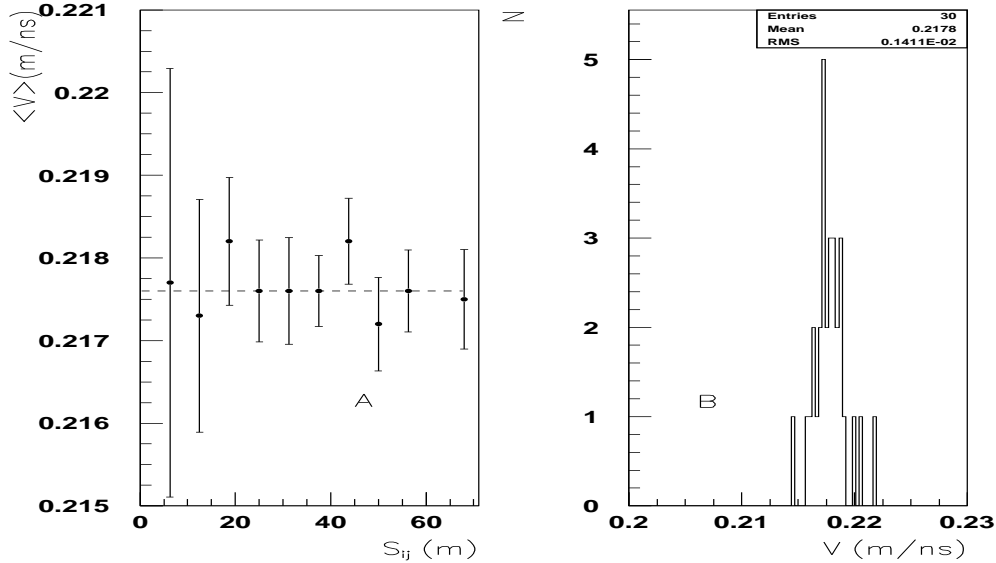,width=15cm,height=9cm}}
\caption[1]{\small The results of $in$ $situ$ measurements of 
light signal velocity /9/.}
\end{figure}   
This result is comparable with predicted value  2.1919$\cdot10^{10}$cm/s 
(see table and Fig.4), the distinction does not exceed one standard deviation
($\sim0.7\%$ of the measured value) and it is possible 
to conclude with high confidence level that 
the experimental light signal velocity value agrees with the 
calculated value of light 
group velocity and does not agree with the phase velocity 
value. To verify an 
influence of salinity, pressure and 
temperature on velocity the precision of measurements has to be increased.\

  Next more extensive and more precise measurements have 
  been done with the experimental 
 string developed near NT-200 site to test new deep underwater techniques /14/. 
 Light Beacons supplied short flashes with fixed photon intensity 
 were used as light sources in these measurements.
 UV($\lambda$ =370 $\pm$ 6 nm),blue ($ \lambda$ =470 $\pm$ 11 nm) 
 and green
 ($\lambda$ =520 $\pm$ 17 nm) LEDs were
  used in Light Beacon. LED flashes
 were detected by four standard NT-200 Channels attached to experimental string.
 The measurements of the time
 intervals between trigger signals of LED drivers and time responses 
 of Chennels, the time delay in electronics and cables (obtained in the
 laboratory) and the distances between light source and Channels (31m, 34m, 
 58m, 64m) were used to get the light signal velocity. In contrast with first 
 experiment there were no ability to estimate the influence of the  
 time walk effect, nevertheless the results : (2.148$\pm0.010)\cdot10^{10}$ cm/s 
 ($\lambda=370\pm 6$ nm); ($2.193\pm0.009)\cdot10^{10}$ cm/s ($\lambda=470\pm$11 nm);
 ($2.206 \pm 0.009)\cdot10^{10}$ cm/s ($\lambda=520 \pm$ 0.017 nm) are in the excellent agreement
 with the predicted values which are 2.1516$\cdot10^{10}$ cm/s, 
 2.1920$\cdot10^{10}$ cm/s and 2.2029$\cdot10^{10}$ cm/s 
 correspondingly (see also  Fig4.).
\begin{center}
{\bf 5. Summary and conclusion}
\end{center}

The measured properties /10$\div$12/ of the water refractive index 
as functions of light wavelength($\lambda$), salinity(S), pressure(P) 
and temperature(t) have been employed in the performed analysis as fully 
as possible.\

The structure of dispersion formula for water was applied which is 
similar with dispersion formula structure for an optical glass /10 /. This
structure was used to describe sucsesfully (see formula (5) and Fig.1) 
the experimental 
refractive index dependence on the light wavelength in the 
interval (200$\div$1250)nm for distilled 
water with t=20$^0$C and P=P$_{atm}$ /10$\div$12/. The polynomial using/14/
gives true result only in the interval (300$\div$600)nm (see Fig.1) in which 
Cherenkov light contributes to the signal of a muon.\

Analysis of the extensive table data /12/ 
shows that formula (5) may be also successfully used (see fig. 1)
for water with other 
properties (S, P, t) just by means adding some 
correction $\bigtriangleup$n$_{SPt}$(S, P, t) which does not 
depend on $\lambda$.\

The assumption is used that equality 
$\bigtriangleup$n$_{SPt}$=$\bigtriangleup$n$_{St}$+$\bigtriangleup$n$_P$ is
sufficiently true in the considered case, and these corrections are
computed for any water states using expressions (11$\div$15).\

To simplify  calculations of the light group velocity the suitable 
formula (16) has been selected which quite satisfactorily 
describe the computed group velocity values in the light wavelength
interval 200$\div$1250 nm (see Fig. 4).\

The comparison of the group velocity value calculated for Baikal 
Neutrino Telescope site with the results of $in$ $situ$ measurement /9, 14/
demonstrates good agreement, the distinctions do not exceed one standard
deviation ($\sim 0.7\%$ of measured value) /9/ and 0.4 standard deviation /14/
while the difference between 
calculated phase velocity value 
and the measured light velocity amounts to four standard 
deviations. Moreover the results /14/ permit to confirm 
a validity of $\bigtriangleup$v$_{gr}$(S, P, t) estimations (see Fig.4).\

The obtained parametric expressions (5), (6), (10$\div$16) could be used  
to calculate a velocity of monochromatic light signal 
for calibration of TOF system by means of a power laser 
lighting through water. These expressions are also suitable for using in
simulation of events detection  and in reconstruction procedure for existed
Underwater Neutrino Telescope projects and for future
1$km^3$ Neutrino Telescopes /13/.\
\begin{center}
{\bf Acknowledgements}
\end{center}
The author is grateful to Dr. E. Bugaev from INR for reading this paper 
and useful remarks.  

\end{document}